\documentclass[twocolumn,amsmath,amssymb,prd]{revtex4-1}
\usepackage{amsmath, amssymb, amsfonts}
\usepackage{graphicx}
\usepackage{float}
\usepackage{physics}
\usepackage[utf8]{inputenc}
\usepackage[colorlinks]{hyperref}
\usepackage{multirow}
\usepackage{natbib}
\hypersetup{linkcolor=black}
\hypersetup{citecolor=black}

\begin{document}
\title{The effect of stout smearing on the phase diagram from \\ multiparameter reweigthing in lattice QCD}

\author{M. Giordano}
\author{K. Kapas}
\author{S. D. Katz}
\author{D. Nogradi}
\author{A. Pasztor}

\affiliation{ELTE E{\"o}tv{\"o}s Lor{\'a}nd University, Institute for Theoretical Physics, P{\'a}zm{\'a}ny P{\'e}ter s. 1/A, H-1117, Budapest, Hungary}

\begin{abstract}
The phase diagram and the location of the critical endpoint (CEP) of lattice QCD with unimproved staggered fermions on a $N_t=4$ lattice was determined fifteen years ago with the multiparameter reweighting method by studying Fisher zeros. We first reproduce the old result with an exact algorithm (not known at the time) and with statistics larger by an order of magnitude. As an extension of the old analysis we introduce stout smearing in the fermion action in order to reduce the finite lattice spacing effects. First we show that increasing the smearing parameter $\rho$ the crossover at $\mu = 0$ gets weaker, i.e., the leading Fisher zero gets farther away from the real axis. Furthermore as the chemical potential is increased the overlap problem gets severe sooner than in the unimproved case, therefore shrinking the range of applicability of the method. Nevertheless certain qualitative features remain, even after introducing the smearing. Namely, at small chemical potentials the Fisher zeros first get farther away from the real axis and later at around $a\mu _q = 0.1 - 0.15$ they start to get closer, i.e., the crossover first gets weaker and later stronger as a function of $\mu$. However, because of the more severe overlap problem 
the CEP is out of reach with the smeared action.
\end{abstract}

\maketitle

\section{Introduction}
One of the most interesting open problems in the study of QCD is determining its phase diagram in the temperature ($T$)-baryonic chemical potential ($\mu _B=3\mu _q=3\mu$) plane. It is established that near $\mu=0$ there is an analytic crossover \cite{Aoki:2006we,Bhattacharya:2014ara} at $T=150-160$ MeV \cite{Aoki:2006br,Aoki:2009sc,Borsanyi:2010bp,Bazavov:2011nk}. It is conjectured  that at higher chemical potential this crossover gets stronger and above a point it turns into a first order phase transition. The endpoint of the line of first order transitions is called the critical endpoint. 

The direct study of the phase diagram is not possible due to the sign problem; however, several extrapolation methods are available including the analytic continuation from imaginary chemical potential \cite{deForcrand:2002hgr,DElia:2002tig,DElia:2009pdy,Cea:2014xva,Bonati:2014kpa,Cea:2015cya,Bonati:2015bha,Bellwied:2015rza,DElia:2016jqh,Gunther:2016vcp,Alba:2017mqu,Vovchenko:2017xad,Bonati:2018nut,Borsanyi:2018grb,Bellwied:2019pxh,Borsanyi:2020fev}, the Taylor expansion method \cite{Allton:2002zi,Allton:2005gk,Gavai:2008zr,Basak:2009uv,Borsanyi:2011sw,Borsanyi:2012cr,Bellwied:2015lba,Ding:2015fca,Bazavov:2017dus,Bazavov:2018mes,Giordano:2019slo,Bazavov:2020bjn}, 
and the reweighting method \cite{Hasenfratz:1991ax,Barbour:1997ej,Fodor:2001au,Fodor:2001pe,Fodor:2004nz,Csikor:2004ik}. All of these methods are based on the fact that the sign problem is absent at $\mu=0$ or at purely imaginary chemical potential. One can therefore use the standard simulation techniques there and then try to reconstruct the theory at real $\mu>0$. 

Among these methods reweighting has no other systematic error besides the overlap problem, thus in principle it can lead to the correct results with infinite statistics. 
With this method, the location of the critical endpoint has been 
determined for $N_t=4$ lattice QCD with an unimproved staggered action \cite{Fodor:2001pe,Fodor:2004nz}, but there 
are still no results closer to the continuum limit. This is due to reweighting methods getting prohibitively expensive for large lattices. 
One possible way to reduce the finite lattice spacing effects without increasing the lattice size is to improve the UV behavior of the action, e.g., by introducing stout smearing \cite{Morningstar:2003gk} in the fermion action. 

In this paper we first reproduce the old $N_t=4$ results with the unimproved action with an exact algorithm\cite{Clark:2006wp} (not known at the time of Refs.~\cite{Fodor:2001pe,Fodor:2004nz}) and much larger statistics. Second, we use one step of stout smearing with several values of the $\rho$ smearing parameter and study the behavior of the phase diagram and the severity of the overlap problem at nonzero chemical potential as a function of $\rho$. Our simulations use $N_f=2+1$ rooted staggered fermions with physical quark masses. The ambiguity of rooting at finite $\mu$ was first discussed in detail in Ref.~\cite{Golterman:2006rw} and also recently emphasized in Ref.~\cite{Giordano:2019gev} where the so-called geometric matching method was introduced to deal with the analyticity issues of the rooted staggered determinant. Here we will use the geometric matching method and compare it to standard rooting.

\section{Multiparameter reweighting}
At finite $\mu _B$ the fermionic determinant $\mathrm{det} \mathrm{M}$ is generally complex and importance sampling methods cannot be applied. This problem can be circumvented with the reweighting technique. The main idea 
is to rewrite the grand canonical partition function as:
\begin{equation}
\begin{split}
Z&=\int \mathcal{D} \mathrm{U} \ \mathrm{det} \mathrm{M} (\mu) e ^ { -S_g (\beta ) }= \\
 &=\int \mathcal{D} \mathrm{U} \ \mathrm{det} \mathrm{M} (0) e ^ { -S_g (\beta_0 ) }  w(\mu,\beta) \rm,
\end{split}
\label{eq:1}
\end{equation}
where for a fixed gauge configuration
\begin{equation}
\label{eq:w}
 w(\mu,\beta)=\frac{\mathrm{det} \mathrm{M} (\mu)}{\mathrm{det} \mathrm{M} (0)}  e ^ { -S_g (\beta)+S_g (\beta _0 ) } \ \rm.
\end{equation}
The configurations can now be generated with importance sampling methods since $\mathrm{det} \mathrm{M} (0)$ is a positive real number. The weight $w(\mu,\beta)$ is instead treated as an observable. Even though Eq.~\eqref{eq:1} is exact, in practice its application is limited to small enough values of the chemical potential and of the volume. This is due to the fact that the tails of the distribution of the weights $w(\mu,\beta)$ are hard to sample. This is known as the overlap problem and is exponentially severe in the volume. Reweighting in $\beta$ was introduced to decrease the severity of this problem, with $\beta _0$ chosen to be close to the value of the gauge coupling at the crossover at zero chemical potential in order to have a more variable ensemble \cite{Fodor:2001au,Fodor:2001pe}.

The analytical form of the staggered fermion determinant on a fixed gauge configuration can be expressed at arbitrary $\mu$ with the eigenvalues $\lambda _i$ of the so called reduced matrix \cite{Hasenfratz:1991ax}:
\begin{equation}
\mathrm{det}\mathrm{M}(a\mu)=e ^ { -3N_s ^3 N_t a \mu } \prod _{i=1} ^{6N_s ^3} \left[ e ^ {N_t a \mu} - \lambda _i \right] \rm,
\label{eq:det1}
\end{equation}
where $N_s$ and $N_t$ are the linear spatial and temporal size of the lattice in lattice units, and the $\lambda _i$ depend on the gauge fields. This determinant describes four flavors of quarks. In order to describe $N_f<4$ the rooting trick is invoked, i.e., $\mathrm{det} \mathrm{M}$ is replaced by $(\mathrm{det} \mathrm{M})^{N_f /4}$. In particular, for $N_f=2$ the complex square root is taken. This introduces a sign ambiguity of the rooted determinant. At $\mu>0$ one usually chooses the complex root that continuously connects to the positive root at $\mu=0$ for each factor in Eq.~\eqref{eq:det1}, i.e., 
the first factor in the Eq.~\eqref{eq:w} becomes:
\begin{equation}
\left. \sqrt{\frac{ \mathrm{det}\mathrm{M}(a\mu)}{ \mathrm{det}\mathrm{M}(0)}} \right| _{\mathrm{rew}} \equiv e ^  { - \frac{3}{2} N_s ^3 N_t a \mu } \prod _{i=1} ^{6N_s ^3} \sqrt{ \frac{ e ^ {N_t a \mu} - \lambda_i  }{ 1 - \lambda_i }} \ \rm.
\label{eq:det}
\end{equation}
In Ref.~\cite{Giordano:2019gev} a new approach, dubbed geometric matching, has been proposed in order to remove the branch point singularities in Eq.~\eqref{eq:det} configuration by configuration. This makes the partition function an entire function of $\mu$ already at finite lattice spacing, which is not the case 
when one uses Eq.~\eqref{eq:det}. The new definition of the corresponding reweighting factor reads:
\begin{equation}
\left. \sqrt{\frac{ \mathrm{det}\mathrm{M}(a\mu)}{ \mathrm{det}\mathrm{M}(0)}} \right| _{\mathrm{P}} \equiv
e ^ { - \frac{3}{2} N_s ^3 N_t a \mu } \prod _{i=1} ^{3N_s ^3}\frac{ e ^ {N_t a \mu} - \xi _i  }{ 1 - \xi _i } \ \rm.
    \label{eq:detP}
\end{equation}
where the set of the ${\xi _i}$ are obtained from the set of the ${\lambda _i}$ via geometric matching \cite{Giordano:2019gev}. Essentially, they are the geometric means of the close-lying pairs of $\lambda _i$. In the continuum limit the two methods are expected to give the same results.

Phase transitions can be studied by means of the zeros of the partition function \cite{Lee:1952ig,Fisher:1965}. The zeros as a function of complex $\beta$ are called Fisher zeros \footnote{Note that in Ref.~\cite{Fodor:2004nz} the terminology was different and zeros in complex $\beta$ were called Lee-Yang zeros. We adopt the more conventional terminology in statistical physics where zeros of the partition function in complex $\beta$ are called Fisher zeros, while the term Lee-Yang zeros is used for zeros in complex $\mu$.}. A phase transition is signaled by an accumulation of the zeros near the real axis in the thermodynamic limit. In the reweighting approach one solves:
\begin{equation}
\frac{Z(\beta,\mu)}{Z (\beta_0,\mu=0)}=\expval{w(\mu,\beta)}=0 \ \rm,
\end{equation}
on the complex $\beta$ plane for fixed $\mu$, where $w(\mu,\beta)$ is defined in (\ref{eq:w}).  When using rooted staggered fermions the ratio of determinants in Eq.~\eqref{eq:w} has to be replaced by the
appropriate complex root. In particular, for $N_f=2$ one can use either of Eqs.~\eqref{eq:det} or ~\eqref{eq:detP}.
As the volume $V$ goes to infinity the imaginary part of the zero closest to the real axis goes to a non-zero  constant for a crossover and to zero in case of a true phase transition. At large enough volumes one expects:
\begin{equation}
\mathrm{Im} \beta _F \sim A V^{-b} +\mathrm{Im}\beta _F ^{\infty} \ \rm,
\label{eq:ImBeta}
\end{equation}
where for a first order phase transition $b=1$ and $\mathrm{Im}\beta _F ^{\infty} =0$, for a second order transition $b<1$ and $\mathrm{Im}\beta _F ^{\infty} =0$, while in the absence of phase transition 
$\mathrm{Im}\beta _F ^{\infty} \neq 0$ and $b$ is in general not known. In our infinite volume extrapolations we will use Eq.~\eqref{eq:ImBeta} with $b=1$ in order to determine whether our data is 
consistent with a first order phase transition or not. Our estimate of the critical endpoint will correspond to the first value of $\mu$ where $\mathrm{Im}\beta _F ^{\infty}$ is consistent with zero. 
When $\mathrm{Im}\beta _F ^{\infty}$ is nonzero it can be considered as a measure of the strength of the crossover.

In this paper we studied the effect of stout smearing \cite{Morningstar:2003gk} on the Fisher zeros. Stout smearing is an analytic map that replaces link variables with a suitable average in order to smooth ultraviolet fluctuations. The procedure depends on a parameter $\rho$ which measures the relative weight of the original and the neighboring links in the average. This procedure has been shown to reduce cutoff effects on several observables. Since stout smeared links are used to compute the fermionic determinant, the Fisher zeros become functions of the smearing parameter $\rho$.

\section{Numerical results}
\subsection{Lattice action and algorithm}
We used the plaquette gauge action and $2+1$ flavors of staggered fermions with one step of stout smearing with smearing parameter $\rho$. For $\rho=$\{0, 0.01, 0.02, 0.03\} we have performed simulations near the crossover temperature, which for the different values of $\rho$ correspond to $\beta _0 = $\{5.188, 5.171, 5.154, 5.137\} respectively. The bare quark masses were set to $m_{ud}=0.0092$, $m_s=0.25$. These correspond to the physical Goldstone pion and kaon mass and are the same as in Ref.~\cite{Fodor:2004nz}. In this exploratory study we use the same bare quark masses for all values of $\beta _0$ and $\rho$. We checked that this has at most a $2 \%$ effect on both the scale setting and the pion mass. For each $\rho$ we simulated lattices of size $6^3 \times 4$, $8^3 \times 4$, $10^3 \times 4$, $12^3 \times 4$.
As in Ref. \cite{Fodor:2001pe,Fodor:2004nz} we only introduce a chemical potential for the light quarks, i.e., we use $\mu _s =0$ which corresponds to $\mu _S = \mu _B /3$. 

\begin{figure*}[ht]\centering
\includegraphics[width=8.5cm]{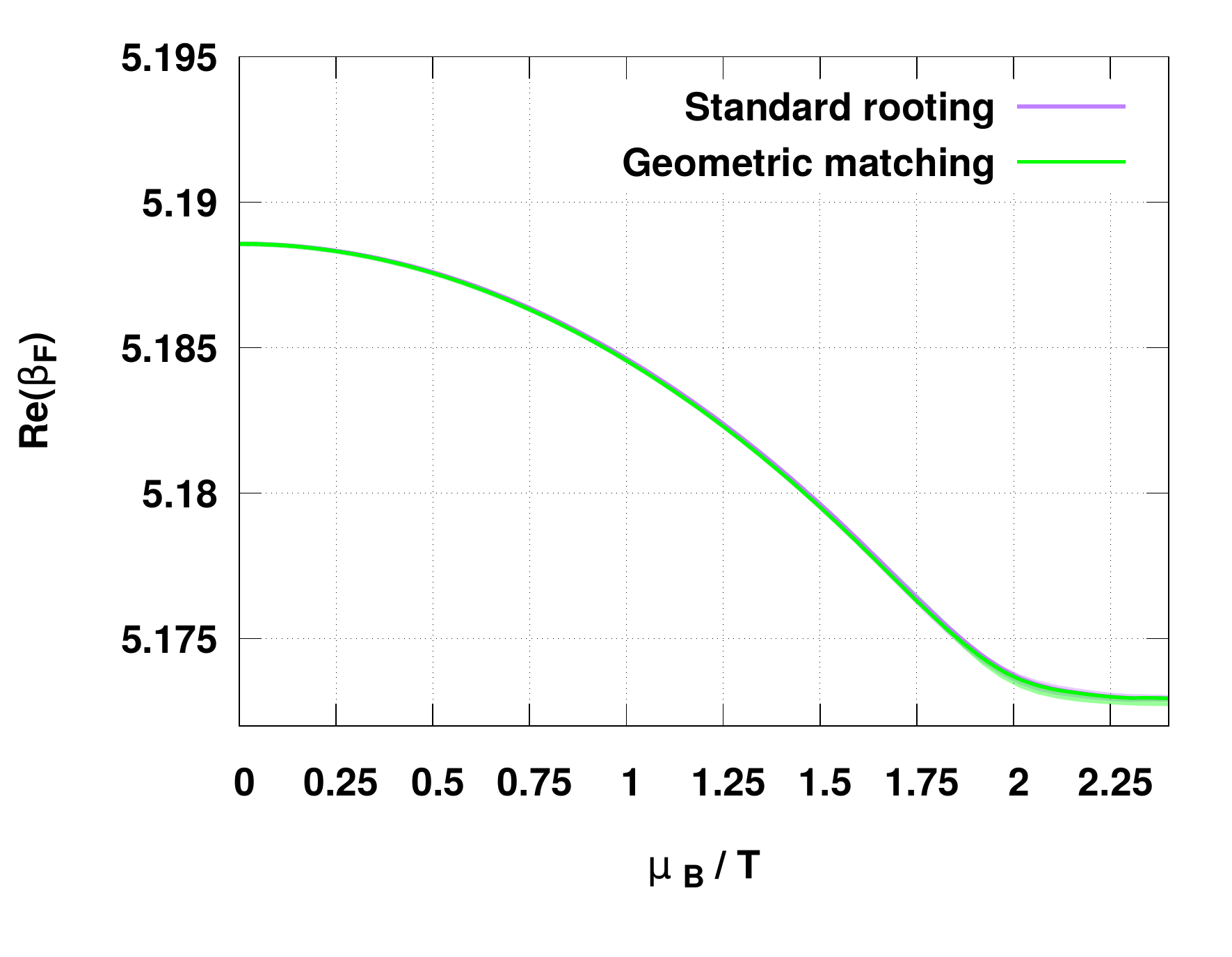}
\includegraphics[width=8.5cm]{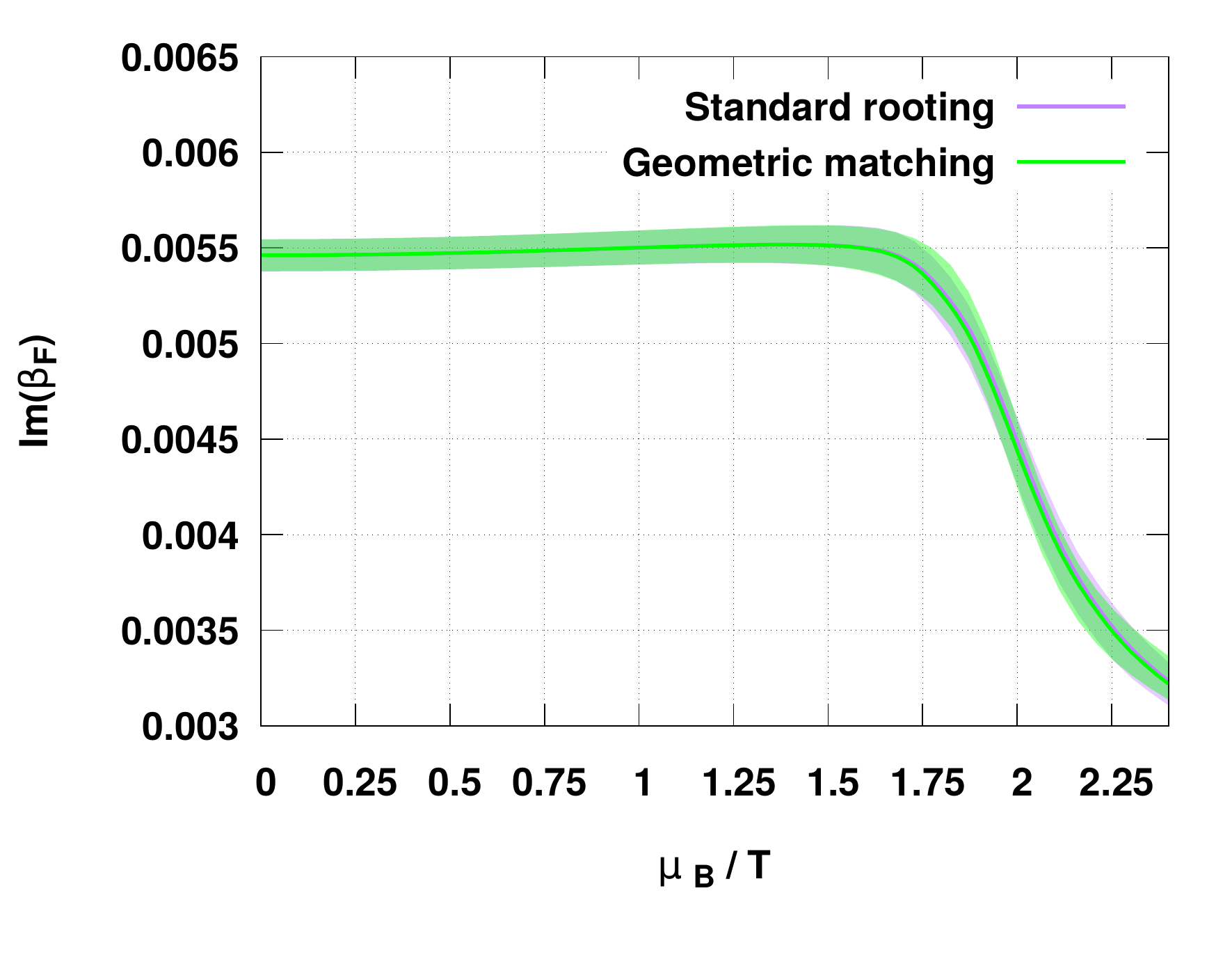}
\caption{The real and imaginary part of the Fisher zero closest to $\beta_0$ as a function of $\mu$ obtained with standard staggered rooting and with geometric matching on a  $12 ^3\cross 4 $ lattice at $\rho=0$. }
\label{fig:comp1}
\end{figure*}
\begin{figure*}[ht]\centering
\includegraphics[width=8.5cm]{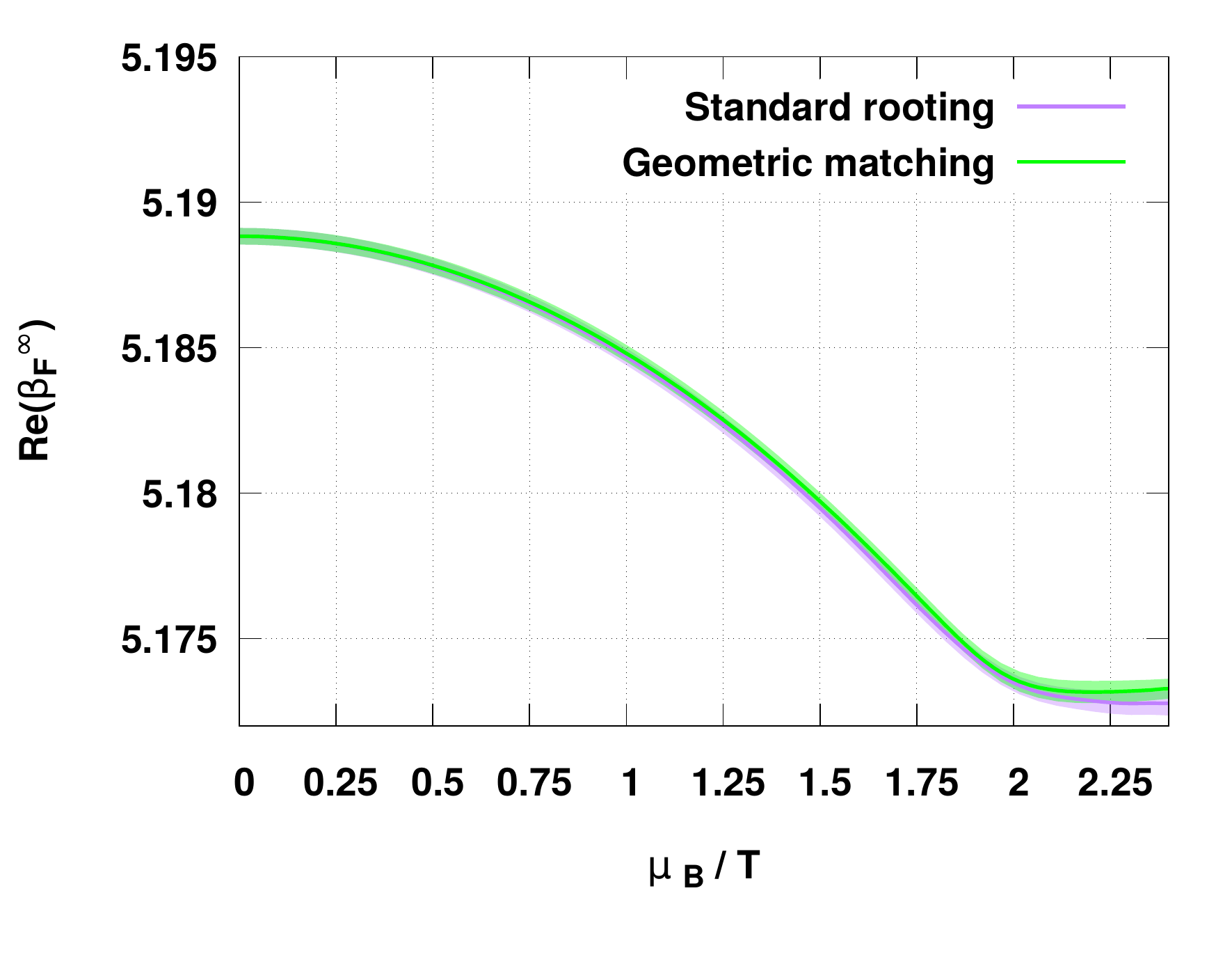}
\includegraphics[width=8.5cm]{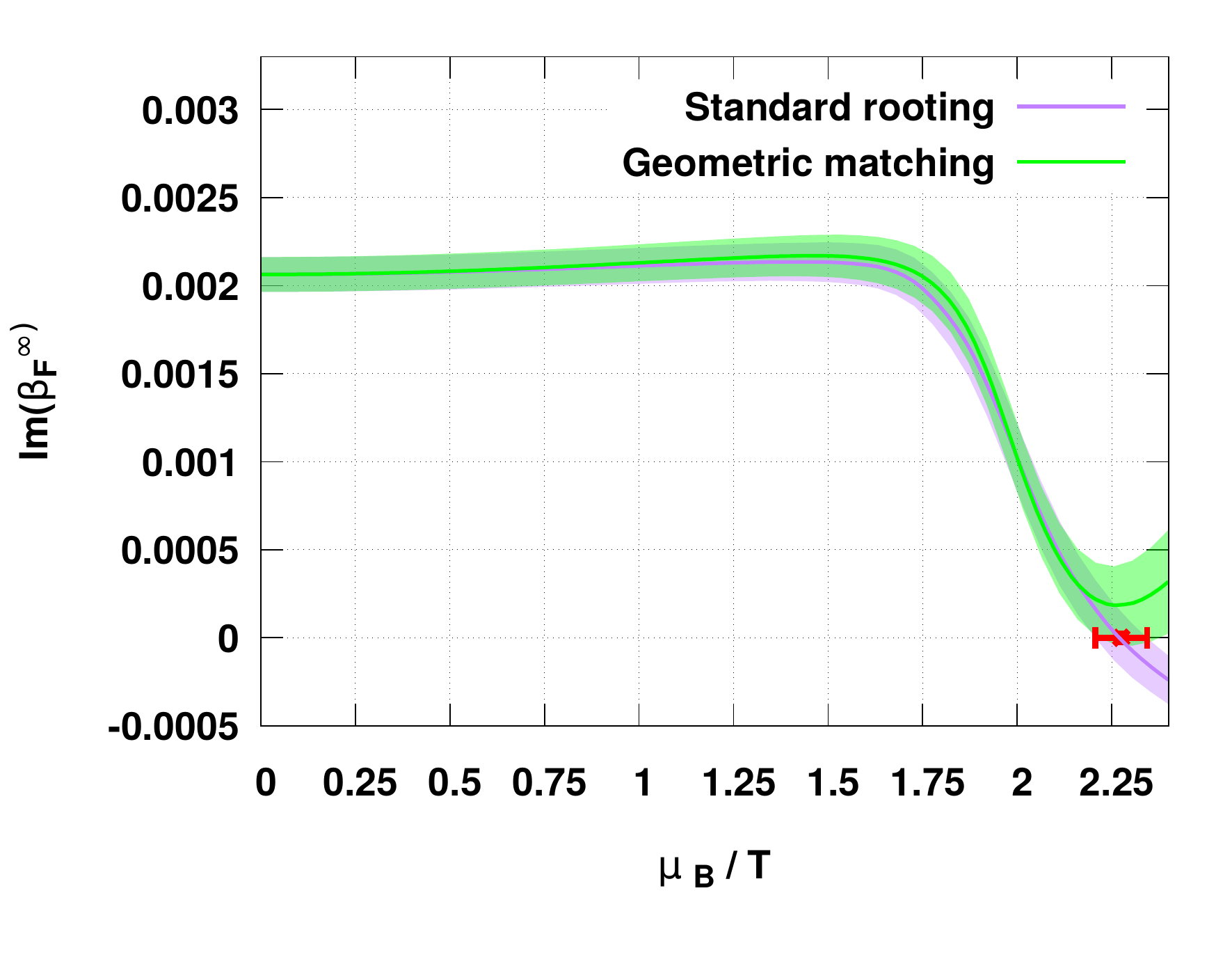}
\caption{The real and imaginary part of the infinite volume extrapolation of the leading Fisher zero via standard rooting and geometric matching at $\rho=0$. 
    The red point with errorbars denotes our estimate of the location of the critical endpoint for $N_t=4$ unimproved staggered fermions. 
    }
\label{fig:comp2}
\end{figure*}

In Ref.~\cite{Fodor:2004nz} the diagonalization required for Eq.~\eqref{eq:det} was carried out on $\sim$ 3000 configurations, which were generated with the R algorithm \cite{Gottlieb:1987mq} with a fixed step size. This is not an exact algorithm, thus it may have uncontrolled systematic errors. 
Here we used the Rational Hybrid Monte Carlo algorithm  \cite{Clark:2006wp}, which is exact. 
Furthermore since the overlap problem  occurs, it is worthwhile to repeat the analysis with much larger statistics. 
For the diagonalization of the reduced matrix, we applied GPU accelerated determinant calculations using the publicly available MAGMA library \cite{tdb10,tnld10,dghklty14}. This allowed us to measure the determinant on an order of magnitude 
more configurations than in Ref. ~\cite{Fodor:2004nz}. The number of configurations for each simulation point is shown in Table \ref{tab:statistics}.

\begin{table}[H]\centering
\begin{tabular}{c c | c c c c}
\hline
\hline
&		 & \multicolumn{4}{c}{$\rho$} \\
&        & $0$ & $0.01$ & $0.02$ & $0.03$ \\
\hline
& 8     & 42215 & 10237 & 10245 & 13196 \\
\multirow{2}{*}{$N_s$} & 10    & 71263 & 22127 & 22363 & 53741 \\
& 12    & 49020 & 26653 & 26861 & 56107 \\
\hline
\end{tabular}
\caption{Number of configurations for each ensemble. Configurations are separated by 10 RHMC trajectories each.}
\label{tab:statistics}
\end{table}

\subsection{Results with the unimproved action}
\label{subsec:comparison}
First we show in Fig.~\ref{fig:comp1} a comparison of the position of the leading Fisher zero, i.e., the one closest to $\beta_0$, 
obtained with the standard rooting and geometric matching for the largest volume in our study at $\rho=0$. 
This assesses the systematics related to the rooting ambiguity. The two methods agree nicely in this case.
\begin{figure*}[ht]\centering
\includegraphics[width=8.5cm]{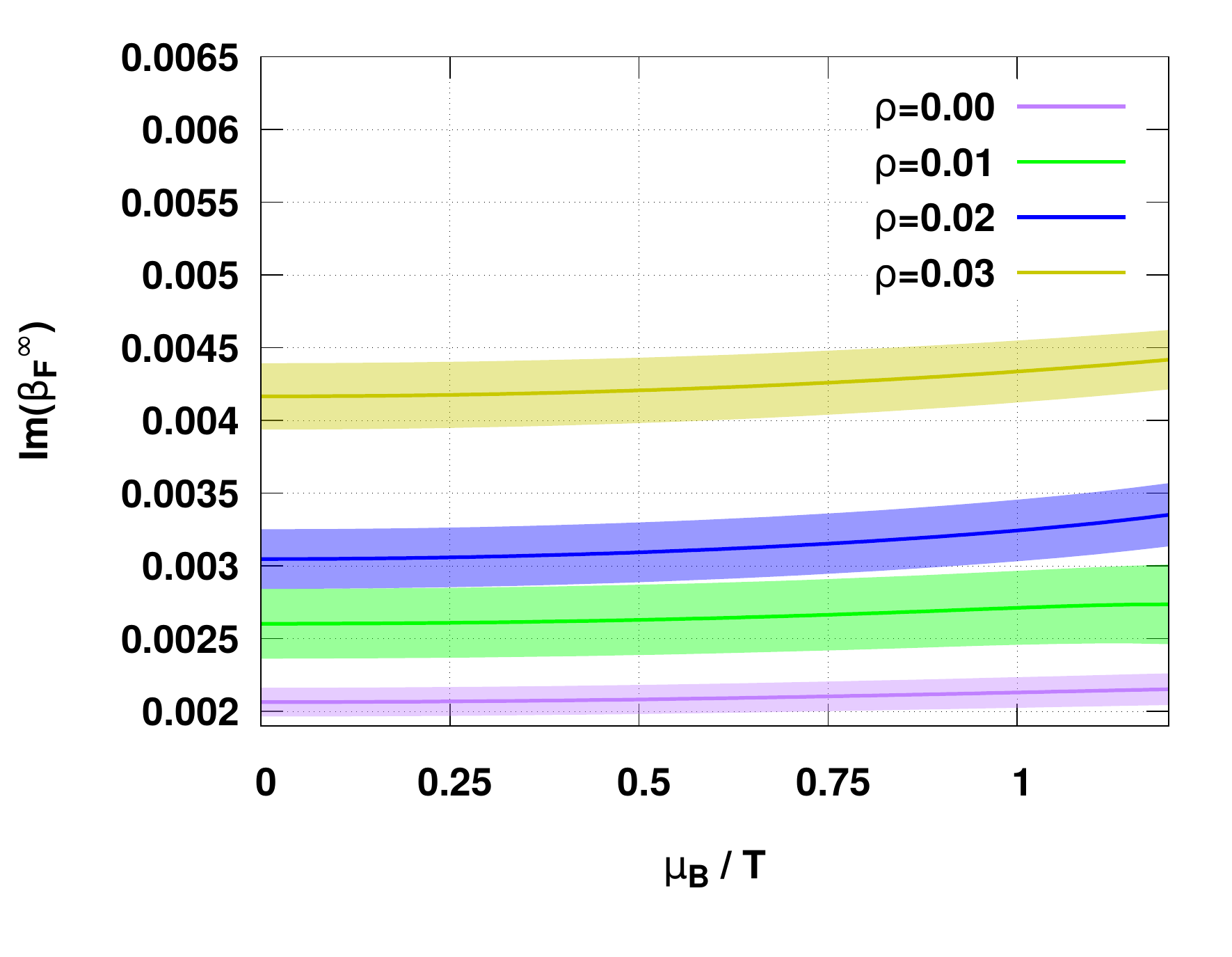}
\includegraphics[width=8.5cm]{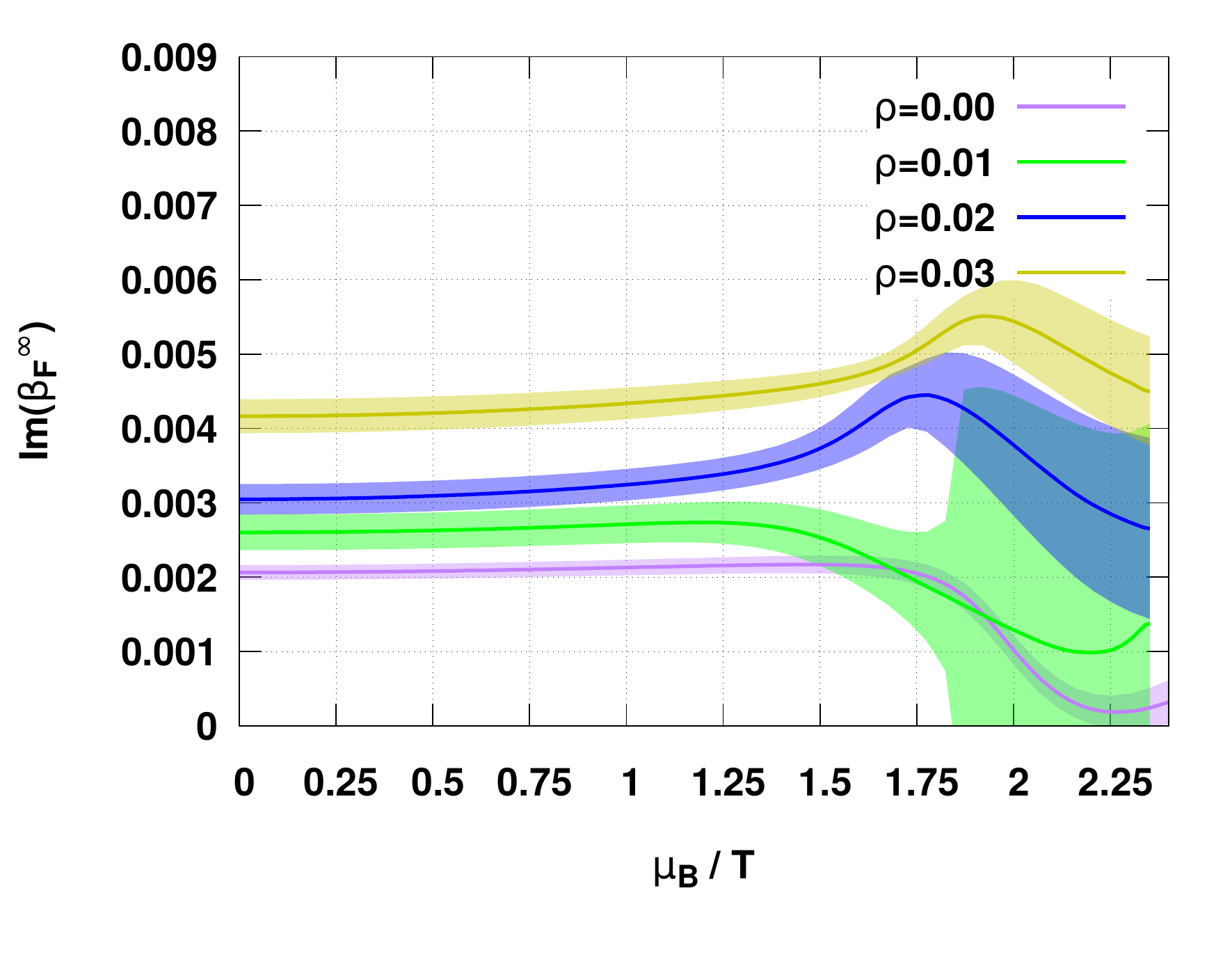}
\caption{The $\mu$ dependence of the imaginary part of the Fisher zeros extrapolated to infinite volume at several values of the smearing parameter $\rho$. The left figure shows only the values close to $\mu=0$. }
\label{fig:infvol_mu0}
\end{figure*}
To estimate the position of the leading Fisher zero we performed an infinite volume extrapolation linear in $1/V$ excluding the smallest volume for both types of rooting. As discussed in the previous section this corresponds to checking whether the data is consistent with a first order phase transition or not. The results are shown in Fig.~\ref{fig:comp2}. As can be seen in Fig.~\ref{fig:comp1} and Fig.~\ref{fig:comp2} the leading Fisher zero as well as our estimate of its infinite volume extrapolation first get farther away from the real axis, i.e., the transition gets weaker at small values of $\mu$. Indeed, on the $12 ^3 \times 4$ lattice the slope of the imaginary part of the leading Fisher zero at $\mu =0$ is positive within  $3\sigma$. At larger $\mu$ the Fisher zero starts to get closer to the real axis, i.e., the crossover strengthens. The value of the chemical potential where the extrapolated imaginary part starts to be consistent with zero is  
\begin{figure*}\centering
\includegraphics[width=4.25cm]{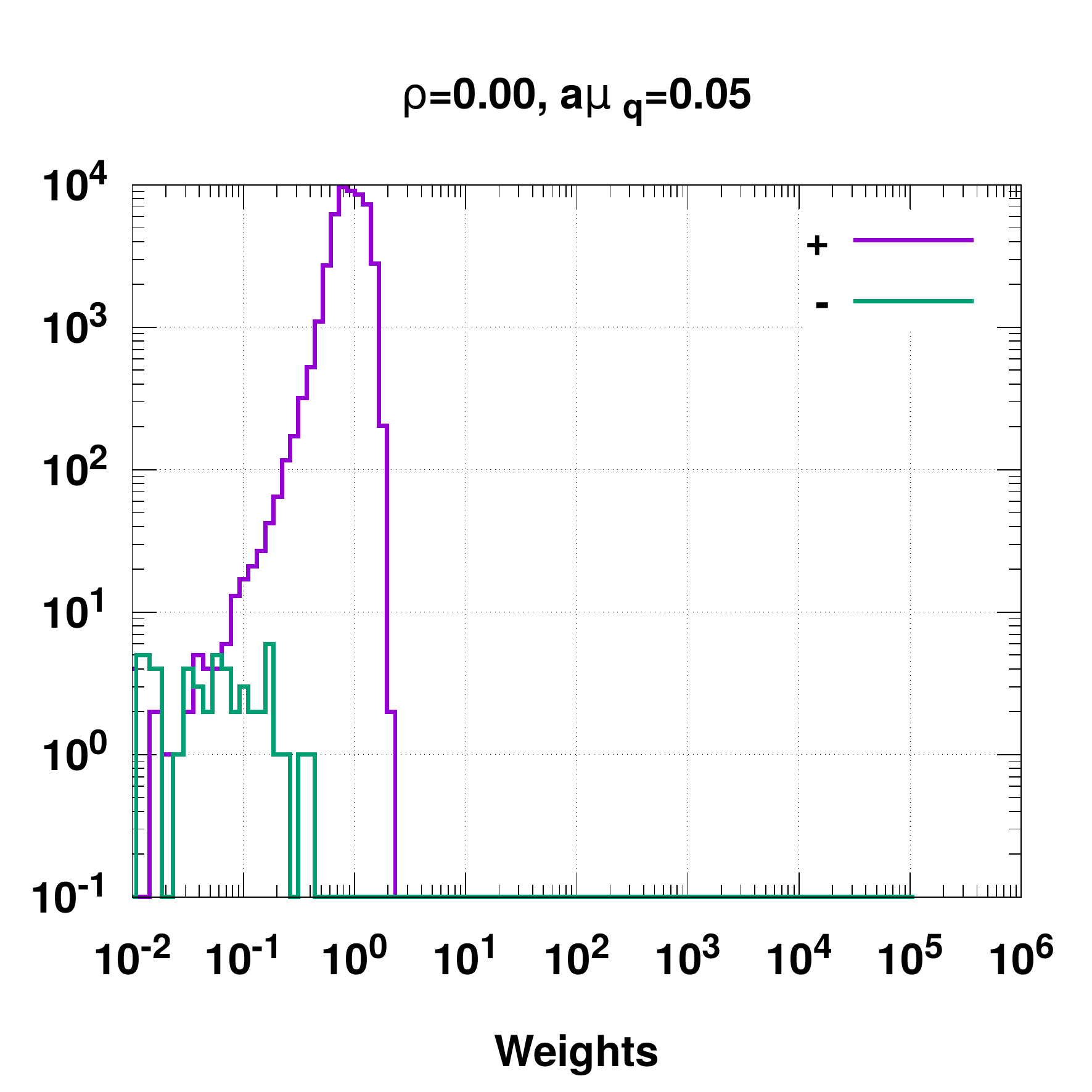}
\includegraphics[width=4.25cm]{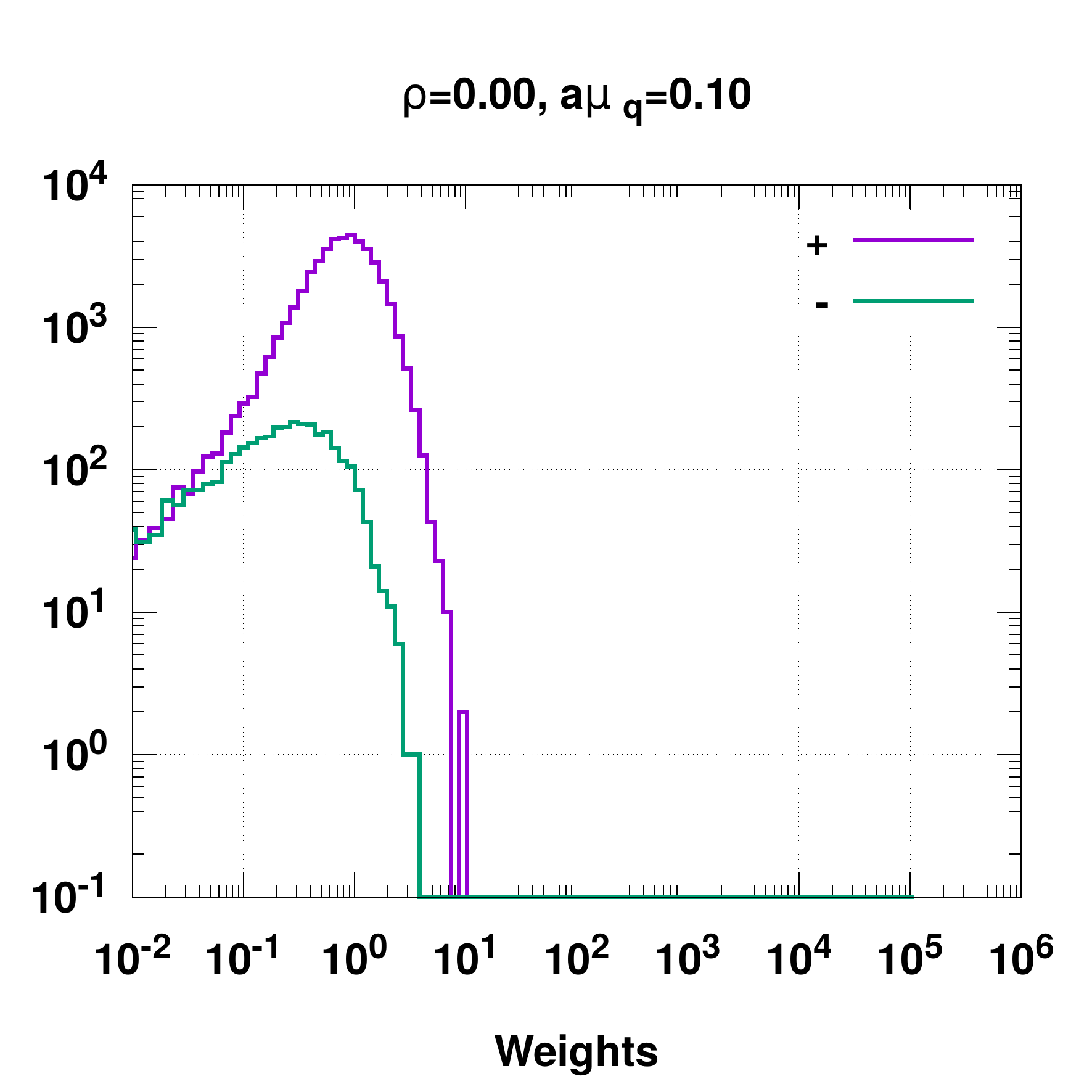}
\includegraphics[width=4.25cm]{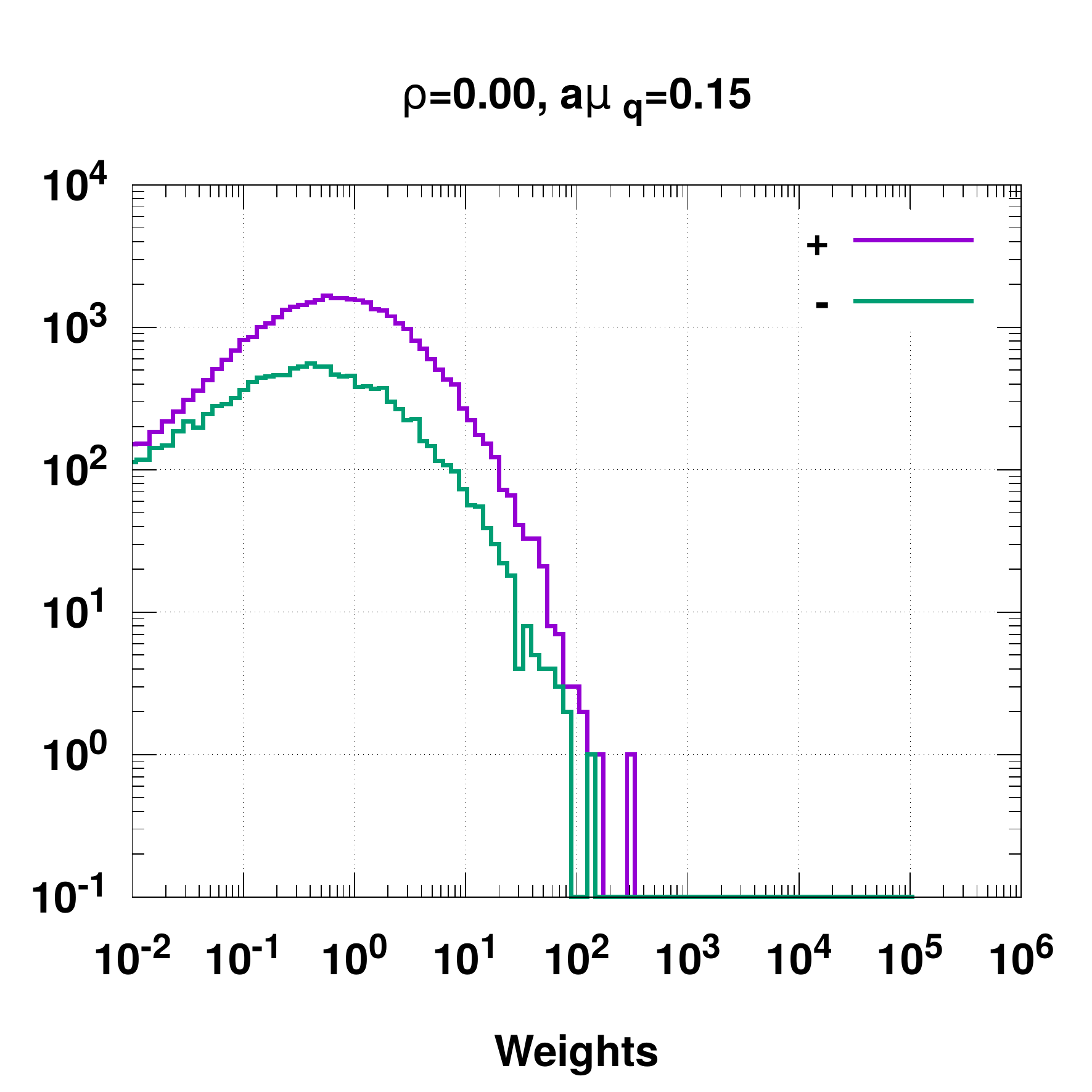}
\includegraphics[width=4.25cm]{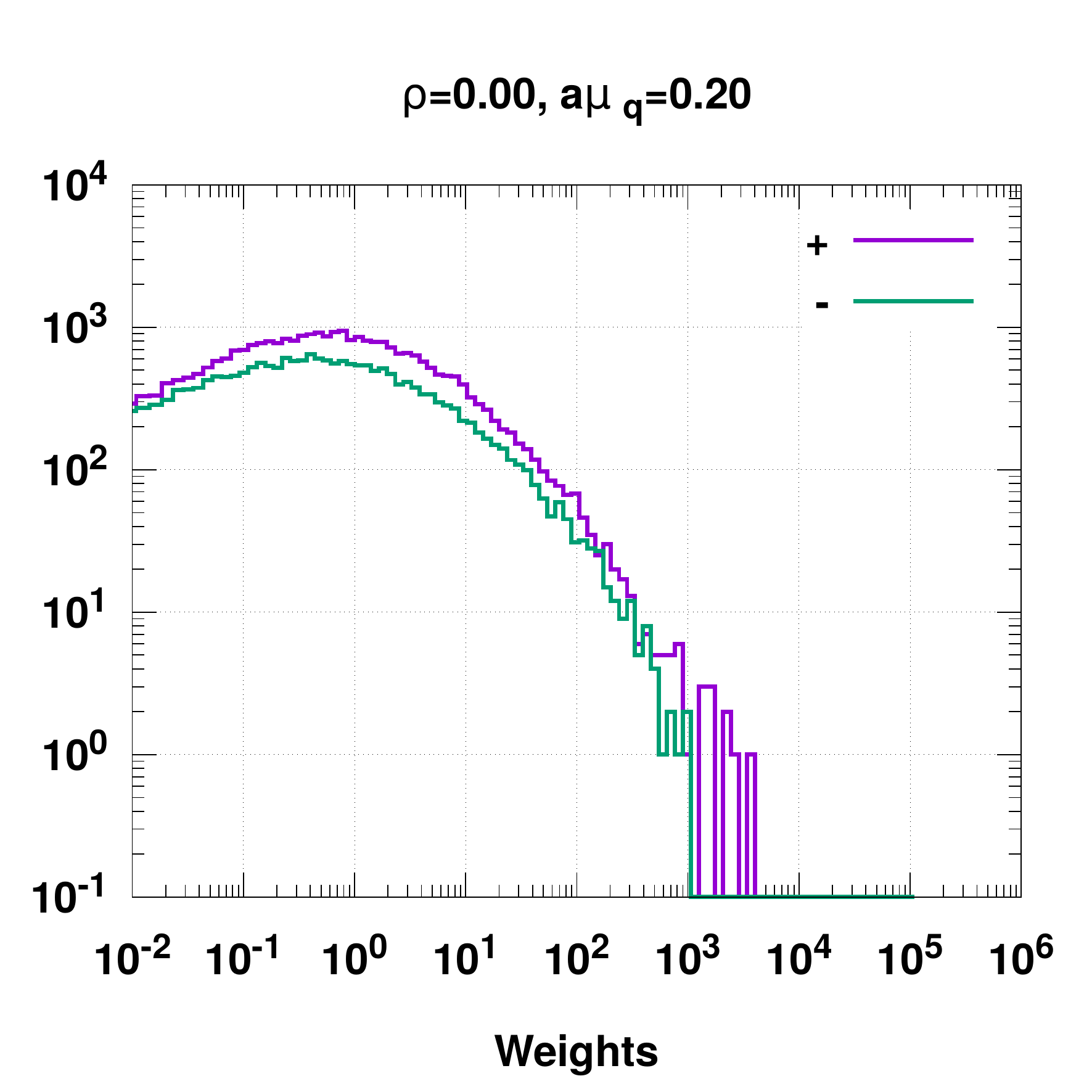}
\includegraphics[width=4.25cm]{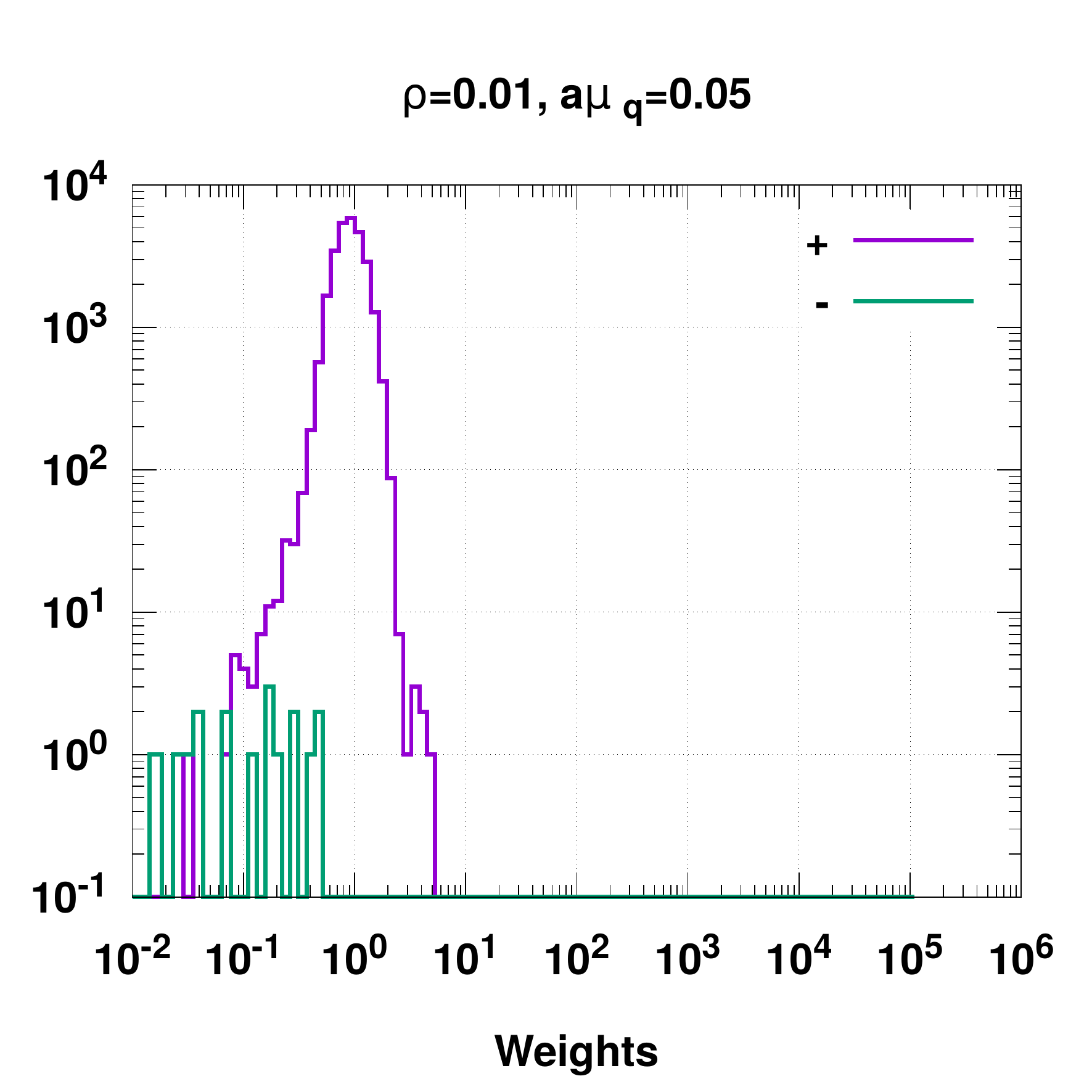}
\includegraphics[width=4.25cm]{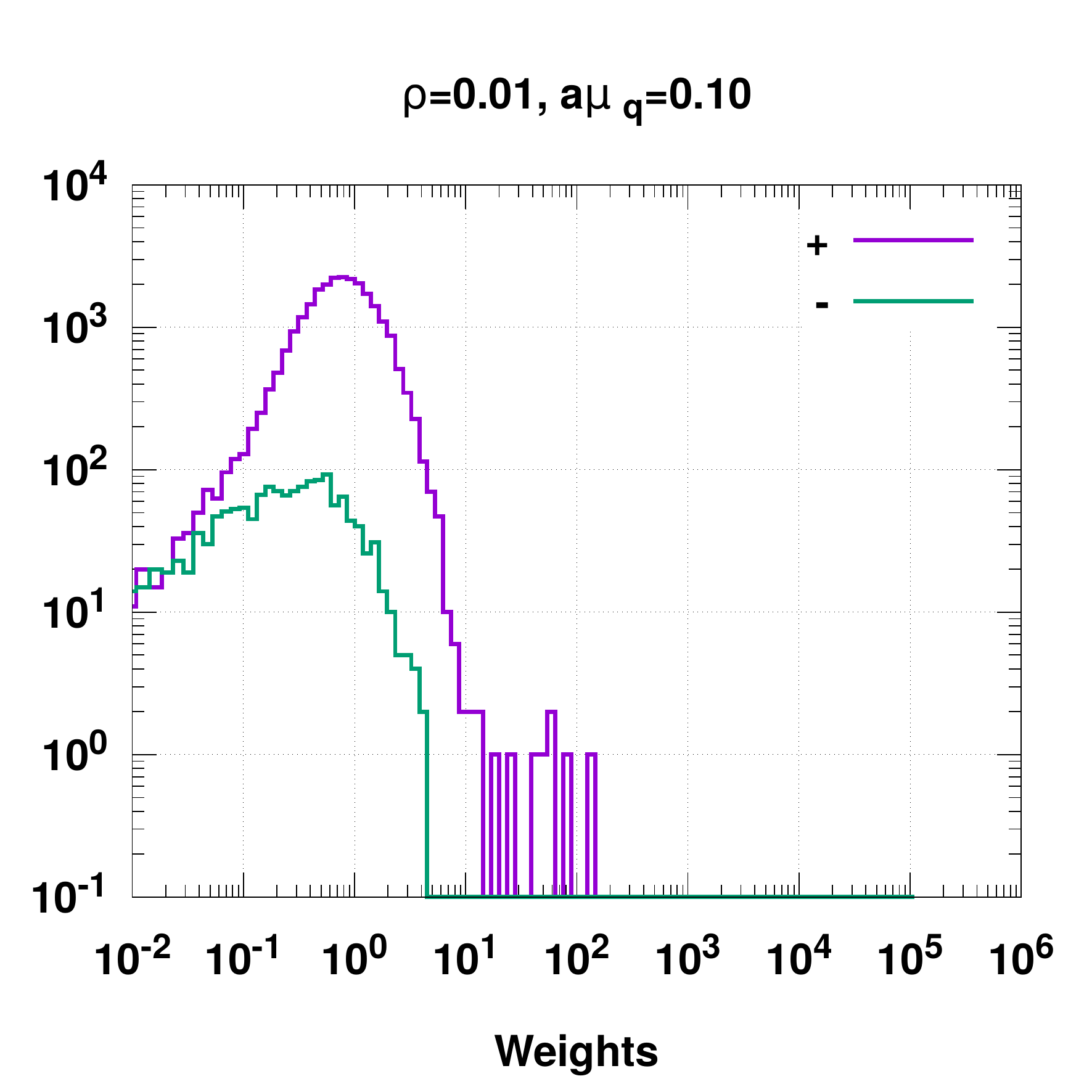}
\includegraphics[width=4.25cm]{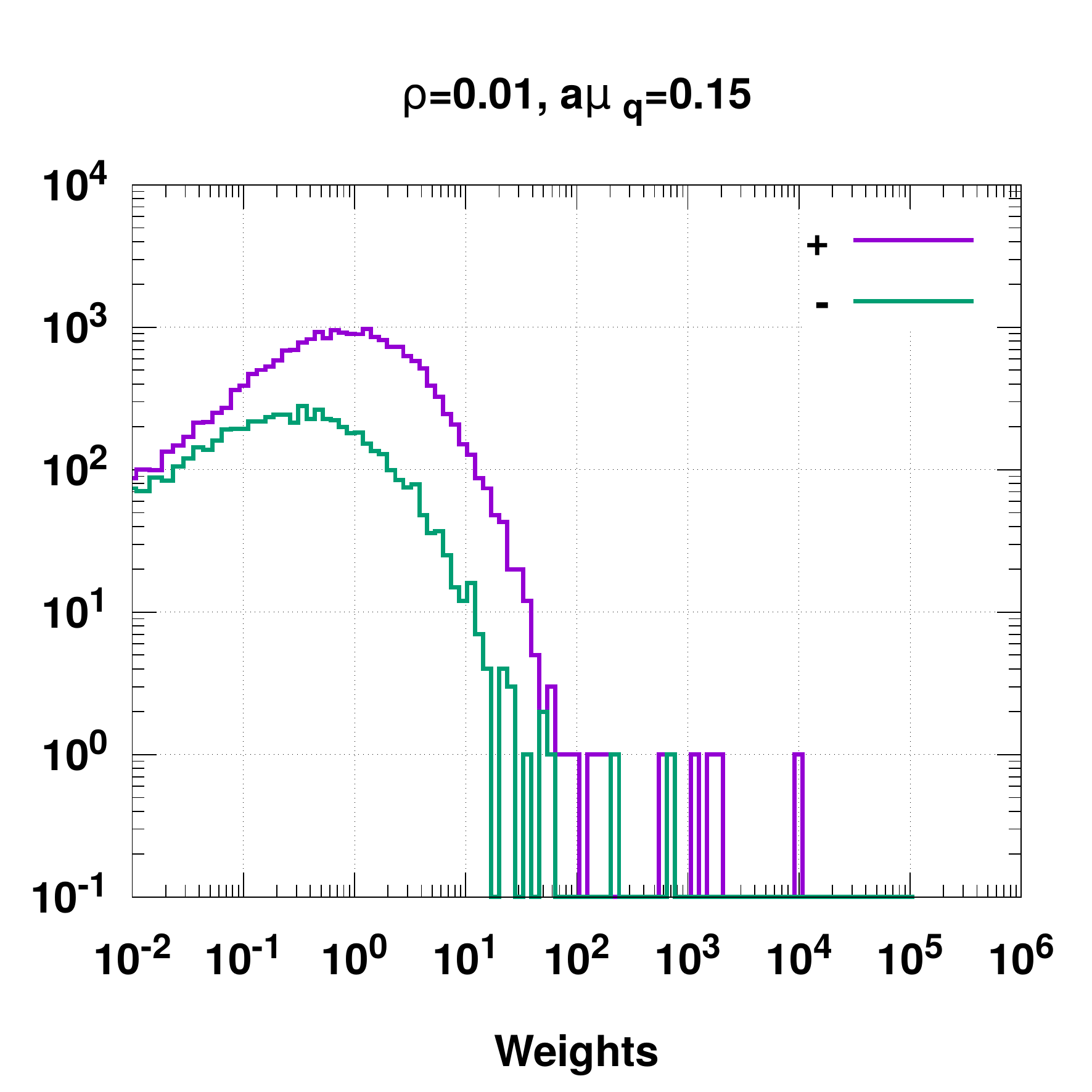}
\includegraphics[width=4.25cm]{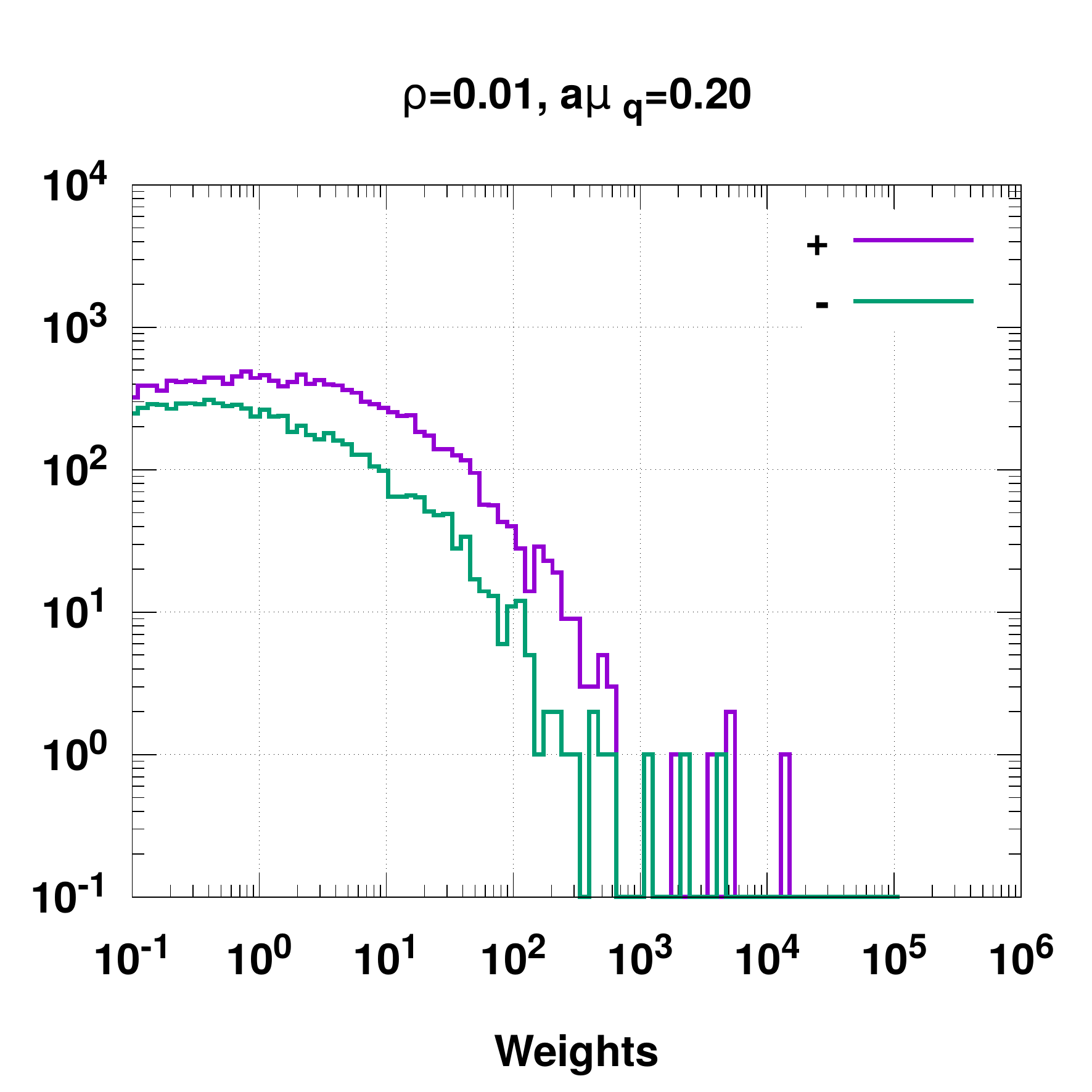}
    \caption{The distribution of the weights $W_i$ of Eq.~\eqref{eq:Wi} for $N _s =12$ for different values of $\rho$ and $\mu$ obtained with geometric matching. 
    The distribution of the positive and negative weights are shown separately on logarithmic scale. At zero smearing there is no outlier configurations even at large $\mu$, however at large $\mu$ and finite smearing configurations appears with extremely large weights. The few outlier configuration appearing indicate a bad sampling of the Monte Carlo integration.
    }
\label{fig:overlap2}
\end{figure*}
\begin{equation}
    \mu _B ^c / T = 2.28 \pm 0.07 \rm{.}
\end{equation}
Here, the errors are purely statistical. 
This is our estimate of the location of the critical endpoint for $N_t=4$ unimproved staggered fermions. Beyond this point estimate the results with the two types of rooting start to disagree, i.e., the ambiguity in the rooting procedure starts to matter. As we will discuss below, in this region the overlap problem is already strong and therefore it is somewhat unclear whether the difference between the two methods is a genuine effect or only an artifact of an incorrect sampling of the tail of the distribution of the weights $w(\mu,\beta)$.
\begin{figure*}[ht]\centering
\includegraphics[width=5.7cm]{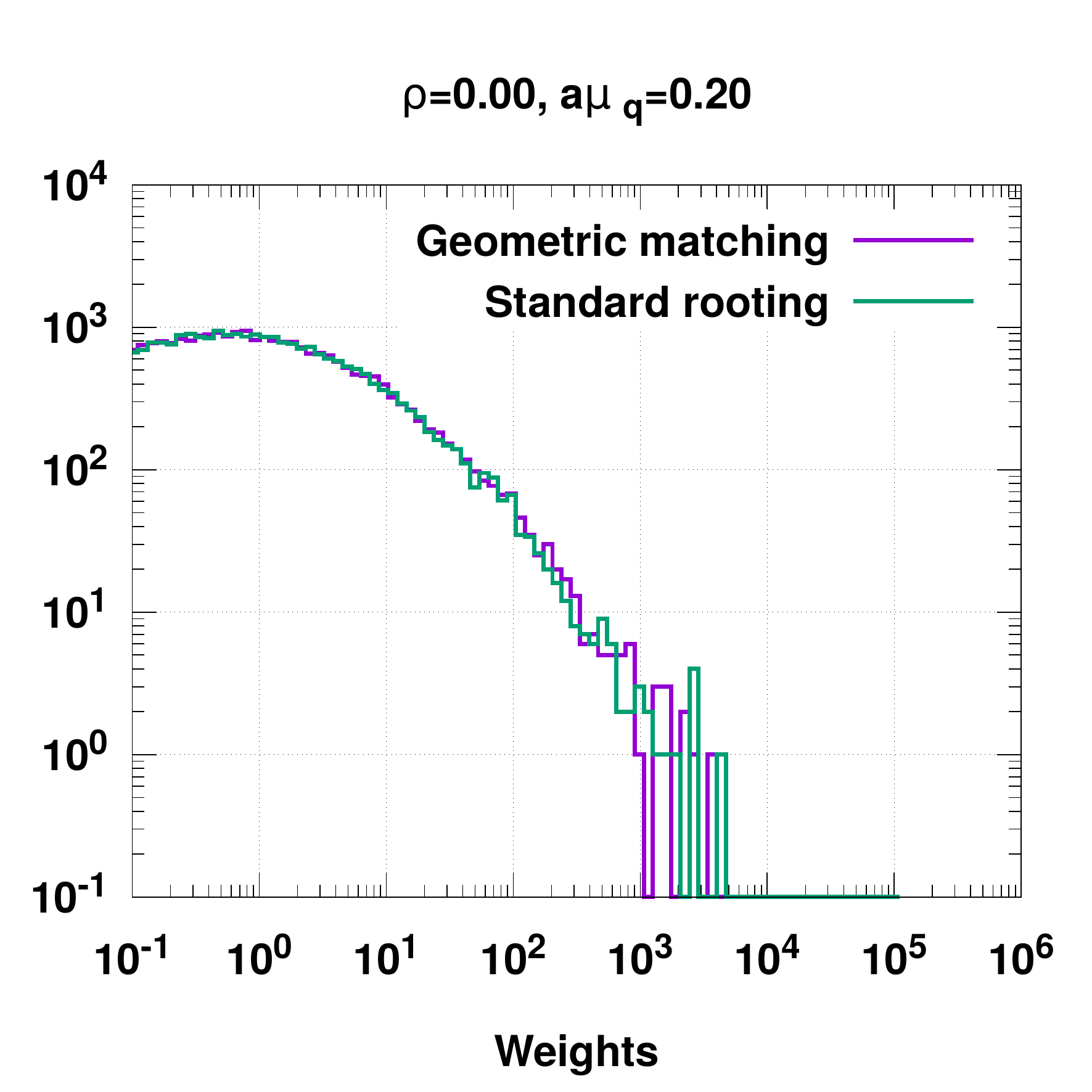}
\includegraphics[width=5.7cm]{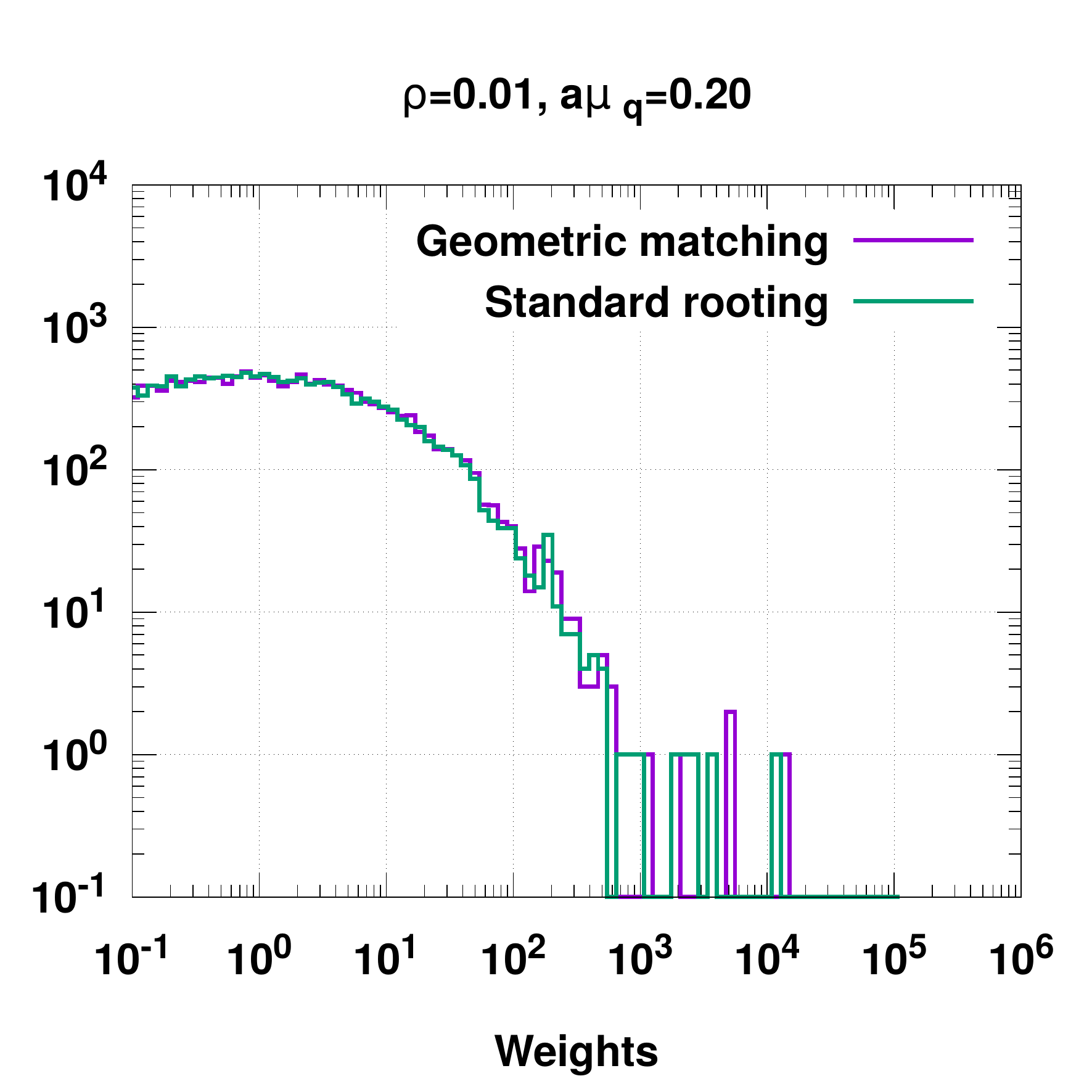}
\includegraphics[width=5.7cm]{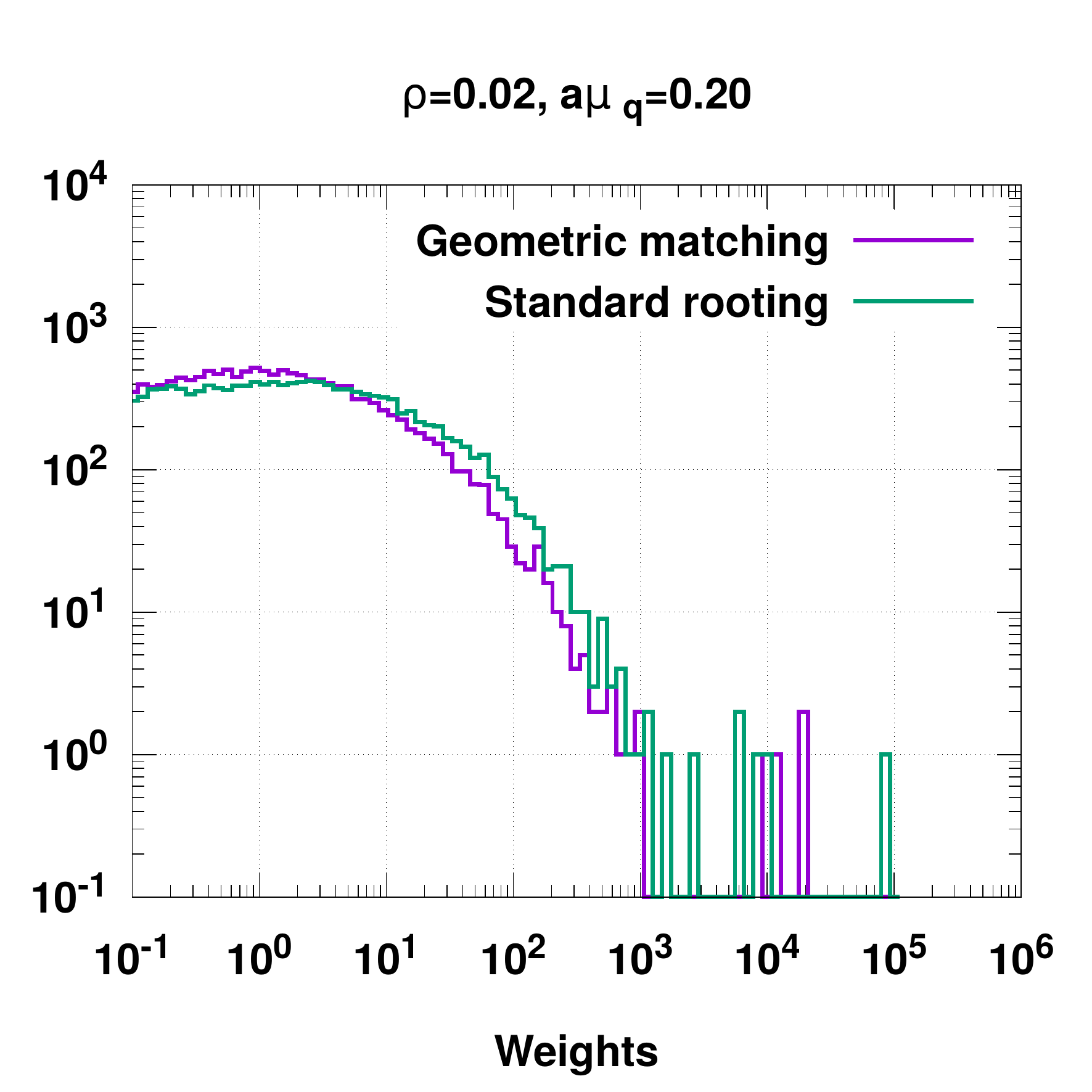}
\caption{Distributions of the positive weights $W_i$ of Eq.~\eqref{eq:Wi} on a $12^3 \cross 4$ lattice at $a\mu=0.2$ for different values of the smearing parameter. 
    }
\label{fig:overlap3}
\end{figure*} 
\subsection{Results with stout smearing}
The infinite volume estimates of the leading Fisher zero position with the different values of the smearing parameter $\rho$ can be seen in Fig.~\ref{fig:infvol_mu0}. Here we only present results obtained with geometric matching. In the left panel we zoom in on the region around $\mu =0$ to emphasize two effects. First, for all values of $\rho$ considered in this work at small chemical potentials the crossover first gets weaker with increasing $\mu$. Second, already at $\mu=0$ the leading Fisher zero gets farther away from the real axis as $\rho$ is increased, i.e., the crossover weakens with increasing $\rho$. This is probably a manifestation of the critical line of the Columbia plot getting farther away from the physical quark masses \cite{deForcrand:2006pv,Varnhorst:2015lea}. In the right panel of Fig.~\ref{fig:infvol_mu0}  we show the full range of chemical potentials. One can see that some qualitative features of the unsmeared case remain: at intermediate values of $\mu$ the transition starts to get stronger. However, the musch larger errors make the determination of the CEP unreliable and we preferred not to attempt it.

As can be seen in the right panel of Fig.~\ref{fig:infvol_mu0} even though the statistics are 
comparable the errorbars at finite smearing are considerable larger than at $\rho=0$. This is 
due to two effects. As we will discuss below the overlap problem gets stronger with finite smearing. It is also known that the strength of the sign problem as measured by the fluctuations of the phase of the determinant is weakened by taste symmetry breaking. For a demonstration of this see Ref.~\cite{Bellwied:2015lba}. Reducing taste symmetry breaking via smearing is then expected to make the sign problem worse. 

To look more closely at the overlap and sign problems we consider the distribution of the real part of the weights $w(\mu,\beta)$ for the case of $\beta= \Re \beta _F (\mu)$, i.e., 
we consider reweighting along the crossover line as defined by the real part of the location of the leading Fisher zero. These weights read:  
\begin{equation}
W_i = \mathcal{N} \Re \left( \sqrt{\frac{\mathrm{det}\mathrm{M}(a\mu)}{\mathrm{det}\mathrm{M}(0)}  } \right) _i \mathrm{e}^{ V \Re (\beta _F - \beta _0) P_i } \rm,
    \label{eq:Wi}
\end{equation}
where $P_i$ is the average plaquette of a single configuration and the rescaling factor $\mathcal{N}$ is chosen such that $\expval{\log W}=0$. In Fig.~\ref{fig:overlap2} we show the distributions of the positive and negative weights, obtained using geometric matching, separately on a logarithmic scale at $\rho=0$ and $0.01$. As $\mu$ increases the two distributions become comparable, signaling the onset of the sign problem. At the same time the support of the histogram broadens and configurations with large reweighting factors become more and more likely. In particular configurations with very large reweighting factors appear but only rarely (``outliers''), indicating that our $\mu=0$ simulations are sampling the tail of the distribution poorly. This signals the onset of the overlap problem. The range spanned by the outliers increases as $\mu$ is increased, indicating that the overlap problem gets worse at larger chemical potentials. Moreover, it also increases when smearing is introduced, again indicating a worsening of the overlap problem. 
 
In Fig.~\ref{fig:overlap3} we compare the distributions of the positive weights obtained with 
geometric matching and standard rooting at the largest value of $\mu$ considered in this 
paper for different values of the smearing parameter. At $\rho=0$ and $0.01$ the two 
distributions agree nicely except in the outlier region. However since the contribution of 
the outliers is substantial the central value for the Fisher zero can (and does)change 
appreciably, but the statistical errors are also large, since the jacknife samples
without the outliers differ considerably from the central value.
At $\rho=0.02$ the situation is similar. 

It seems then that substantial differences in the Fisher zeros obtained with the two rooting methods appear only when the overlap problem is strong. This is most likely also the case for the unsmeared action at $a\mu _q > 0.2$. 

\section{Summary and conclusion}
The determination of the phase diagram of lattice QCD at finite baryochemical potential, and in particular that of the critical endpoint, are especially difficult due to the infamous sign problem. Bypassing this problem by extrapolating from $\mu_B=0$, where it is absent, to finite $\mu_B$ is possible in principle, but hindered by the overlap problem, i.e., the difficulty in sampling correctly the probability distribution of the relevant observables. Such a problem is exponentially severe in the volume of the system. Despite these difficulties, multiparameter reweighting techniques were employed successfully in Refs.~\cite{Fodor:2001pe,Fodor:2004nz} to determine the CEP using unimproved staggered fermions on lattices of fixed temporal size $N_t=4$ with physical quark masses. In this paper we have confirmed this result using an exact algorithm~\cite{Clark:2006wp} not available at the time (so reducing systematic errors), much higher statistics (so reducing the statistical error), and a new definition~\cite{Giordano:2019gev} of the rooted fermion determinant which improves the analyticity properties of the partition function. The approach used here is based on the study of the large-volume limit of the Fisher zeros of the partition function at finite $\mu_B$ using multiparameter reweighting techniques. We have found for the critical endpoint $\mu _B ^c / T = 2.28 \pm 0.07$, a result compatible with that of Ref.~\cite{Fodor:2004nz}.

Extrapolation of this result to the continuum is obviously difficult, in particular due to the need of bigger lattices that would make the overlap problem worse. A possible way to reduce UV effects and bring the system effectively closer to the continuum is to employ stout smearing~\cite{Morningstar:2003gk} on the gauge links. In this paper we have studied the effect of one step of smearing with small smearing parameter on the Fisher zeros of the partition function at finite $\mu_B$. Unfortunately smearing turns out to make the overlap problem worse, making it appear sooner, i.e., at lower values of $\mu_B$, and making the determination of the critical endpoint unreliable. This is probably related to the critical line of the Columbia plot moving away from the physical point when smearing is introduced \cite{deForcrand:2006pv,Varnhorst:2015lea}. In fact, for the unimproved action the leading Fisher zero(i.e., the one closest to the real $\beta=\beta_0$ where the simulations are performed) is relatively close to the real axis at $\mu_B=0$, thus allowing a sufficiently accurate sampling of the relevant configurations for the finite-$\mu_B$ physics using simulations at $\mu_B=0$. This reflects the presence of a genuine phase transition at $\mu_B=0$ for quark masses smaller but not much smaller than the physical ones \cite{deForcrand:2006pv}. On the other hand, it is known that smearing pushes the critical values of the masses away from the physical point \cite{Varnhorst:2015lea}. This is expected to lead to a weakening of the transition, which should reflect into the leading Fisher zero moving away from the real axis. This is precisely what we observed in our study: even for small values of the smearing parameter the imaginary part of the leading Fisher zero grows appreciably already at $\mu_B=0$. This leads to a poorer sampling at $\mu_B=0$ of the configurations relevant at finite $\mu_B$, and an earlier onset of the overlap problem. This is
particularly evident if one looks at the distribution of the reweighting factors used to reconstruct the finite-$\mu_B$ theory, focusing on the presence of outliers with large reweighting factors. These are a symptom of the poor sampling of the tails of the distribution of the weights, and become more and more relevant as $\mu_B$ increases. It turns out that smearing makes the problem worse, with outliers appearing already at smaller $\mu_B$.

As a side analysis, we have compared the Fisher zeros obtained using the recently introduced 
geometric matching method \cite{Giordano:2019gev}, with those obtained with the standard rooting. While the two methods are expected to 
lead to the same continuum limit, finite lattice spacing effects could be very different. Our results with the two methods are compatible 
whenever the overlap problem is absent or mild, indicating that they have comparable cut-off effects at zero and small $\mu_B$, at least for the lattices 
considered in this work. Incompatibility of the two methods only shows up when the overlap problem is severe, and it is not possible to determine whether it 
signals a large difference in their finite lattice spacing effects or whether it has to be ascribed to the overlap problem itself.

In conclusion, the results of this paper indicate that the bottleneck in studying QCD at finite $\mu_B$ with multiparameter reweighting is the overlap and not
the sign problem. Extending our results closer to the continuum limit therefore requires a much better control of the overlap problem. 

\section*{Acknowledgment}
We thank Zolt\'an Fodor for a careful reading of the manuscript.
This work was partially supported by the Hungarian National
Research, Development and Innovation Office - NKFIH grant KKP126769
and by OTKA under the grant OTKA-K-113034. A.P. is supported by
the J\'anos Bolyai Research Scholarship of the Hungarian Academy of
Sciences and by the \'UNKP-19-4 New National Excellence Program of
the Ministry for Innovation and Technology.


\end{document}